\documentclass[a4paper,fleqn,usenatbib]{mnras}
\usepackage{graphicx}
\usepackage{amsmath}
\usepackage{latexsym}
\usepackage{amsfonts}

\title[Reconstruction of inflationary potentials]{The reconstruction of inflationary potentials}

\author[J. Lin et al.]{Jianmang Lin,$^{1}$\thanks{563939360@qq.com} Qing Gao,$^{2,1}$\thanks{gaoqing01good@163.com} and Yungui Gong,$^{1}$\thanks{yggong@mail.hust.edu.cn} \\
$^1$School of Physics, Huazhong University of Science and Technology, Wuhan 430074, China\\
$^2$School of Physical Science and Technology, Southwest University, Chongqing 400715, China}

\begin{document}
\maketitle

\begin{abstract}
The observational data on the anisotropy of the cosmic microwave background constraints the scalar spectral tilt $n_s$ and
the tensor to scalar ratio $r$ which depend on the first and second derivatives of the inflaton potential. The information
can be used to reconstruct the inflaton potential in the polynomial form up to some orders. However, for some classes
of potentials, $n_s$ and $r$ behave as $n_s(N)$ and $r(N)$ universally in terms of the number of e-folds $N$.
The universal behaviour of $n_s(N)$ can be used
to reconstruct a class of inflaton potentials. By parametrizing one of the parameters $n_s(N)$,
$\epsilon(N)$ and $\phi(N)$, and fitting the parameters in the models to the observational data, we
obtain the constraints on the parameters and reconstruct the classes
of the inflationary models which include the chaotic inflation, T-model, hilltop inflation,
s-dual inflation, natural inflation and $R^2$ inflation.

\end{abstract}



\section{Introduction}
The quantum fluctuation during inflation seeds the large-scale structure and imprints the information of early Universe in
the anisotropies of the cosmic microwave background (CMBR). For the slow-roll inflation with a single scalar field, the scalar \citep{Mukhanov:1981xt} and tensor \citep{starobinsky79}
power spectra can be parametrized by the slow-roll parameters \citep{Stewart:1993bc}.
In particular, the scalar spectral tilt $n_s$, the amplitude of the scalar spectrum $A_s$ and the tensor to scalar ratio $r$
can be constrained by the observational data on CMBR. For an inflationary model, we can
calculate the slow-roll parameters and the observables $n_s$ and $r$, and compare the results with the observations.
However, there are lots of inflationary models with different potentials \citep{Martin:2014vha} and it is not an easy task to compare all the models
and constrain the model parameters in the models, although we may use Bayesian evidence to select best models \citep{Martin:2013nzq}.
Recently, the temperature and polarization measurements on the CMBR by the Planck survey gave
the results $n_s=0.9645\pm 0.0049$ (68\% CL), $\ln(10^{10} A_s)=3.094\pm 0.034$ (68\% CL)
and $r_{0.002}<0.11$ (95\% CL) \citep{Adam:2015rua,Ade:2015lrj}.
If we take the number of e-folds before the end of inflation at the horizon exit $N=60$, then the measured scalar spectral tilt
can be approximated as $n_s-1\approx -2/N$, and $r$ can be described
as $r\sim N^{-p}$ with $p\ge 1$. The small value of $r$ alleviates
the problem imposed by the Lyth bound \citep{Choudhury:2013iaa,Gao:2014yra,Choudhury:2014sua,Gao:2014pca}.
Since the running of the scalar spectral index is a second-order effect, it is expected to be in the order of $10^{-3}$
for slow-roll inflationary models \citep{Gong:2014cqa}.
The running of the scalar spectral index is constrained to be $d n_s/d\ln k=-0.0057\pm 0.0071$ (68\% CL) \citep{Ade:2015lrj}.

For chaotic inflation with the power-law potential $V(\phi)=V_0\phi^p$ \citep{linde83}, we get $n_s-1=-(p+2)/(2N)$ with $N$ being the number
of e-folds before the end of inflation at the horizon exit. For the hilltop models with the potential $V(\phi)=V_0(1-\mu \phi^p)$ \citep{Boubekeur:2005zm},
we get $n_s-1=-2(p-1)/[(p-2)N]$  and $r\sim 0$. For the Starobinsky model \citep{starobinskyfr}, we get $n_s-1=-2/N$ and $r=12/N^2$.
For a class of inflationary models with non-minimal coupling to gravity \citep{Kallosh:2013tua}, it was found that $n_s-1=-2/N$ and $r=12/N^2$.
For natural inflation \citep{Freese:1990rb}, we get $n_s-1=-[1+\exp(-N/f^2)]/[f^2-f^2\exp(-N/f^2)]$ and $r=8\exp(-N/f^2)/[f^2-f^2\exp(-N/f^2)]$.
Therefore, there are some universal behaviours for $n_s$ and $r$ which are consistent with the observations for a large class of inflationary models.
The suggested simple relation  between $n_s$ and $N$ tells us that it will be more easier to compare the models with the observational data if the
slow-roll parameters and the observable $n_s$ can be expressed in terms of $N$. By parametrizing the slow-roll parameter $\epsilon$ as
$\epsilon=\beta/(N+1)^\alpha$, Mukhanov derived the corresponding inflaton potential \citep{Mukhanov:2013tua}.
For the simple inverse power-law form $N^{-p}$, Roset divided some inflationary models into two universal classes with the same $1/N$ behaviour
for $n_s$ and different power-law behaviour for $r$ \citep{Roest:2013fha}. More complicated forms of $\epsilon(N)$ were
also proposed by Garcia-Bellido and Roset \citep{Garcia-Bellido:2014gna}. The parameters $\alpha$ and $\beta$ in the parametrization
were fitted to the observational data in \citep{Barranco:2014ira}. The consequences
of the two fixed points on $n_s$ and $r$ in the parametrization were discussed in \citep{Boubekeur:2014xva}.
In addition to the parametrization of $\epsilon$, the reconstruction of the inflaton potential
from the parametrization of $n_s$ with $n_s-1=-\alpha/N$,
and from the assumption between the amplitude of the power spectrum $-\ln\Delta_R^2$ and $n_s-1$
were also considered \citep{Chiba:2015zpa,Creminelli:2014nqa,Gobbetti:2015cya}. On the other hand, the reconstruction
of the inflaton potential from the power spectra by the method of functional
reconstruction were usually applied \citep{Hodges:1990bf,Copeland:1993jj,Liddle:1994cr,Lidsey:1995np,Peiris:2006ug,Norena:2012rs,Choudhury:2014kma,Ma:2014vua,Myrzakulov:2015fra,Choudhury:2015pqa}.

In this paper, we discuss the reconstruction of the inflaton potential from the parametrizations of $\epsilon(N)$, $\phi(N)$ and $n_s(N)$.
The parameters in the models are fitted to the Planck temperature and polarization data. The paper is organized as follows. In the Section 2,
we review the general relations between $\epsilon(N)$, $\phi(N)$, $n_s(N)$ and $V(N)$ by applying the slow-roll formula. The reconstruction
of the inflaton potential from $n_s(N)$ is presented in Section 3.
The reconstruction of the inflaton potential from $\epsilon(N)$ is presented in Section 4.
The reconstruction of the inflaton potential from $\phi(N)$ is presented in Section 5.
The conclusions are drawn in Section 6.

\section{General Relations}

For the single slow-roll inflation, to the first order of approximation, we have
\begin{equation}
\label{nsapproxeq2}
n_s-1\approx 2\eta-6\epsilon, \quad r=16\epsilon.
\end{equation}
Since
\begin{equation}
\label{nsapproxeq3}
\frac{d\ln\epsilon}{dN}=2\eta-4\epsilon,
\end{equation}
so
\begin{equation}
\label{nsapproxeq4}
n_s-1=-2\epsilon+\frac{d\ln\epsilon}{dN}.
\end{equation}
If we parametrize the slow-roll parameter $\epsilon$ as a function of the number of e-folds before the end of inflation, then we
can derive the parametrization $n_s(N)$ and $r(N)$. Conversely,
if we parametrize $n_s(N)$, we can solve equation (\ref{nsapproxeq4}) to get $\epsilon(N)$, and then $r(N)$.

From the energy conservation of the scalar field, we have
\begin{equation}
\label{sclreq1}
d\ln\rho_\phi+3(1+w_\phi)d\ln a=0.
\end{equation}
Since $\rho_\phi\approx V(\phi)$, $d\ln a =-d N$, and
\begin{equation}
\label{sclreq6}
1+w_\phi=\frac{\dot\phi^2}{\rho_\phi}\approx \frac{\dot\phi^2}{V(\phi)}\approx \frac{2}{3}\epsilon,
\end{equation}
so
\begin{equation}
\label{sclreq2}
\epsilon\approx \frac{1}{2}\frac{d\ln V}{dN}=\frac{1}{2}(\ln V)_{,N}>0,
\end{equation}
and the inflton always rolls down the potential during inflation.
Substituting the above result into equation (\ref{nsapproxeq4}), we get \citep{Chiba:2015zpa}
\begin{equation}
\label{nsapproxeq6}
n_s-1\approx -(\ln V)_{,N}+\left(\ln\frac{V_{,N}}{V}\right)_{,N}=\left(\ln\frac{V_{,N}}{V^2}\right)_{,N}.
\end{equation}
If we have any one of the functions $\epsilon(N)$, $n_s(N)$ and $V(N)$, we can derive the other
functions by using equations. (\ref{nsapproxeq4}), (\ref{sclreq2}) and (\ref{nsapproxeq6}).

To derive the form of the potential $V(\phi)$, we need to find the functional relationship $\phi(N)$ for the scalar field. Note that
\begin{equation}
\label{nphieq1}
d\phi=\pm \sqrt{2\epsilon(N)}dN,
\end{equation}
where the sign $\pm$ depend on the sign of the first derivative of the potential and the scalar
field is normalized by the Planck mass $M_{pl}=(8\pi G)^{-1/2}=1$, so
\begin{equation}
\label{nphieq2}
\phi-\phi_e=\pm \int_0^N \sqrt{2\epsilon(N)}dN.
\end{equation}
Once one of the functions $\epsilon(N)$, $n_s(N)$, $\phi(N)$ and $V(N)$ is known, in principle we can derive
$n_s$, $r$ and the potential $V(\phi)$
by using the relations (\ref{nsapproxeq4}), (\ref{sclreq2}), (\ref{nsapproxeq6}) and (\ref{nphieq1}).

Before we present particular parametrizations, we briefly discuss the effect of second-order corrections.
To the second-order of approximation, we have \citep{Stewart:1993bc,Schwarz:2001vv}
\begin{gather}
\label{4.4.24}
\begin{split}
n_s-1&\approx 2\eta-6\epsilon-\left(\frac{10}{3}+24C\right)\epsilon^2+\frac{2}{3}\eta^2\\
&+(16C-2)\epsilon\eta+\left(\frac{2}{3}-2C\right)\xi,
\end{split}\\
\label{4.4.25}
r\approx 16\epsilon\left[1+\left(4C-\frac{4}{3}\right)\epsilon+\left(\frac{2}{3}-2C\right)\eta\right],\\
\label{rundef}
\frac{dn_s}{dN}\approx -16\epsilon\eta+24\epsilon^2+2\xi,
\end{gather}
where the Euler constant $\gamma\approx 0.577$, $C=\ln2+\gamma-2\approx -0.73$ and
\begin{equation}
\label{xietaeq1}
\xi=\frac{d\eta}{dN}+2\epsilon\eta.
\end{equation}
The observational data requires $n_s-1\sim 2/N$, so $\epsilon$ and $\eta$ are at most in the order of $1/N$, their derivatives
with respect to $N$ have at most the order of $1/N^2$, so $\xi$ is at most in the order of $1/N^2$. Therefore, the running of the
scalar spectral index and the second-order corrections will be in the order of $1/N^2$, we may neglect the second-order
corrections.

\section{The parametrization of the spectral tilt}

We approximate $n_s$ as
\begin{equation}
\label{nsappoxeq1}
n_s-1\approx -\frac{p}{N+\alpha},
\end{equation}
where the constants $p$ and $\alpha$ are both positive, and the constant $\alpha$ accounts for
the contribution from the scalar field $\phi_e$ at the end of inflation.
With this approximation, then $n_s$ is well behaved at the end of inflation when $N=0$.
Note that the functional form of the potential is not affected by $\alpha$.
The observational
results favour $p> 1$, so we consider $p> 1$ only.

For $p>1$, the solution to Equation (\ref{nsapproxeq4}) is
\begin{equation}
\label{epsilonneq1}
\epsilon(N)=\frac{p-1}{2(N+\alpha)+C(N+\alpha)^p},
\end{equation}
where $C\ge 0$ is an integration constant. This is a generalization of the Mukhanov parametrization
$\epsilon(N)=\beta/(N+1)^p$ \citep{Mukhanov:2013tua}.
Since $\epsilon(N=0)\approx 1$, so $2\alpha+C\alpha^p\approx p-1$.
If $C<0$, then $|C|$ must be a very small number to ensure
that $0<\epsilon(N=60)\ll 1$, and its contribution is negligible so that we can take it to be
zero. Therefore, we consider $C\ge 0$ only. Since
\begin{equation}
\label{paramrel1}
C\approx \frac{p-1-2\alpha}{\alpha^p},
\end{equation}
so $C\ge 0$ requires that $p\ge 2\alpha+1$.

Combining Equations (\ref{nsapproxeq3}) and (\ref{epsilonneq1}), we get
\begin{equation}
\label{etaneq1}
\eta(N)=\frac{3(p-1)}{2(N+\alpha)+C(N+\alpha)^p}-\frac{p}{2(N+\alpha)}.
\end{equation}
For $N\gg 1$ and $C\sim 1$, $\epsilon\sim 1/(N+\alpha)^p$ and
$\eta\sim -1/(N+\alpha)$, so the
tensor to scalar ratio $r\sim 1/(N+\alpha)^p$ is small and only $\eta$ contributes to $n_s$.

Either solving Equation (\ref{nsapproxeq6}) with the parametrization (\ref{nsappoxeq1}),
or solving Equation (\ref{sclreq2}) with the solution (\ref{epsilonneq1}),   we get \citep{Chiba:2015zpa}
\begin{equation}
\label{vpareq1}
V(N)=\frac{p-1}{A}\left[\frac{1}{(N+\alpha)^{p-1}}+\frac{C}{2}\right]^{-1},
\end{equation}
where $A>0$ is an integration constant.

Let us consider the special case $C=0$ first. For $C=0$, we get
\begin{equation}
\label{epsilonneq2}
\epsilon(N)=\frac{p-1}{2(N+\alpha) },
\end{equation}
and
\begin{equation}
\label{etaneq2}
\eta(N)=\frac{2p-3}{2(N+\alpha)}.
\end{equation}
Therefore, both $\epsilon$ and $\eta$ contribute to the scalar spectral tilt and
\begin{equation}
\label{nsreq1}
r=\frac{8(p-1)}{p}(1-n_s).
\end{equation}

\begin{figure}
\includegraphics[width=0.4\textwidth]{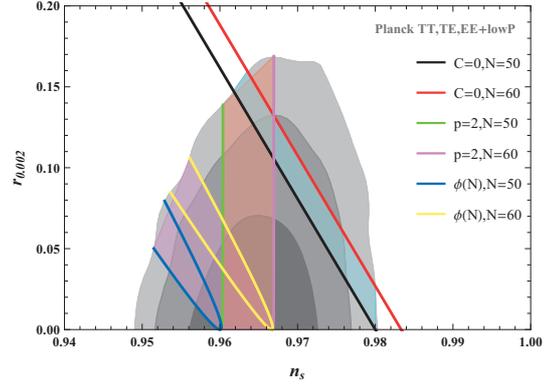}
\caption{The marginalized 68\%, 95\% and 99.8\% CL contours for $n_s$ and $r_{0.002}$ from Planck 2015 data \citep{Ade:2015lrj} and the observational
constraints on the parametrizations \eqref{epsilonneq1} and $\phi(N)=\sigma \ln(\beta N+\gamma)$. The cyan regions are for the
parametrization \eqref{epsilonneq1} with $C=0$, the brown regions are for the parametrization \eqref{epsilonneq1} with $p=2$, and the magenta
areas are for the parametrization $\phi(N)=\sigma \ln(\beta N+\gamma)$.}
\label{nsrpower}
\end{figure}

Substituting Equation (\ref{epsilonneq2}) into Equation (\ref{nphieq2}), we get
\begin{equation}
\label{phineq3}
\phi-\phi_e=\pm 2\sqrt{p-1}(\sqrt{N+\alpha}-\sqrt{\alpha}),
\end{equation}
or
\begin{equation}
\label{phineq4}
\phi(N)=\pm 2\sqrt{(p-1)(N+\alpha)}+\phi_0,
\end{equation}
where $\phi_0$ is an arbitrary integration constant, and $\phi_e=\pm 2\sqrt{(p-1)\alpha}+\phi_0$.
Since $2\alpha\sim p-1$, so $\phi \ga 1$ and this model corresponds to large field inflation. Combining Equations (\ref{phineq4}) and (\ref{vpareq1}),
we get the power-law potential for chaotic inflation \citep{linde83}
\begin{equation}
\label{vphieq2}
\begin{split}
V(\phi)&=\frac{p-1}{A(p-1)^{p-1}}\left(\frac{\phi-\phi_0}{2}\right)^{2(p-1)}\\
&=V_0(\phi-\phi_0)^{2(p-1)}.
\end{split}
\end{equation}
For simplicity, we
apply the Planck 2015 68\% CL constraints on $r$ and $n_s$ to the $r-n_s$ relation \eqref{nsreq1}  \citep{Ade:2015lrj},
and we find that no $p$ satisfies the 68\% CL constraints as shown in Fig. \ref{nsrpower}.
At about the 99.8\% CL level, $1.00 < p\leq1.93$ and $0 < \Delta\phi\leq 12.39$
for $N=50$, $1.20\leq p\leq2.21$ and $6.65 \leq\Delta\phi\leq 15.42$ for $N=60$.
From the above analysis, we see that the power-law potential is disfavoured by the observations at the 68\% CL,
and the filed excursion for the inflaton is super-Planckian. For $N=50$, $p\sim 1$ is marginally consistent
with the observational constraint at the 99.8\% CL, it is possible that the field excursion of the inflaton
is sub-Planckian and the tensor to scalar ratio $r$ is close to zero.

Next we consider the special case with $p=2$. For this case, we get
\begin{gather}
\label{phineq5}
\phi(N)=\pm \frac{2\sqrt{2}}{\sqrt{C}}{\rm arcsinh}\sqrt{C(N+\alpha)/2}+\phi_0,\\
\label{phieneq3}
\phi_e=\pm \frac{2\sqrt{2}}{\sqrt{C}}{\rm arcsinh}\sqrt{C\alpha/2}+\phi_0,
\end{gather}
where $\phi_0$ is an arbitrary integration constant and $C=(1-2\alpha)/\alpha^2>0$. Combining Equations (\ref{phineq5}) and (\ref{vpareq1}),
we get the corresponding T-model potential \citep{Kallosh:2013hoa},
\begin{equation}
\label{vphieq3}
\begin{split}
V(\phi)&=\frac{2}{AC}\tanh^2\left[\frac{\sqrt{C}}{2\sqrt{2}}(\phi-\phi_0)\right]\\
&=V_0\tanh^2[\gamma(\phi-\phi_0)],
\end{split}
\end{equation}
where $\gamma=\sqrt{C/8}$. If $C\ll (N+\alpha)^{-1}$ or $\alpha\rightarrow 1/2$, then the potential reduces to the quadratic potential. From the
discussion on the power-law potential, we know that the inflaton is a large field
and it is disfavoured by the observations at the 68\% CL level. If $C>1$ or $\alpha\ll 1$,
then the potential becomes
\begin{equation}
\label{vphieq4}
\begin{split}
V(\phi)&\approx V_0\left\{1-2\exp\left[-\sqrt{C/2}\,(\phi-\phi_0)\right]\right\}^2\\
&\approx  V_0\left\{1-4\exp\left[-\sqrt{C/2}\,(\phi-\phi_0)\right]\right\}.
\end{split}
\end{equation}
The potential includes the models with $\alpha$-attractors \citep{Kallosh:2013yoa} and the
Starobinsky model \citep{starobinskyfr} when $C=4/3$ or $\alpha=(\sqrt{21}-3)/4$.
By fitting Equations (\ref{nsappoxeq1}) and (\ref{epsilonneq1}) to the Planck 2015 data, for $N=50$,
we get $\alpha<0.4993$ and $0<\Delta\phi\leq 13.90$ at the 99.8\% CL, for $N=60$, we get $\alpha<0.5009$
and $0<\Delta\phi\leq 15.49$ at the 99.8\% CL. The results are shown in Fig. \ref{nsrpower}.
Here we extend the integration constant $C$ to the region of $C<0$, and we verify the conclusion that $|C|$ is very
small if $C<0$ as discussed above. The above results also tell us that this model can be either
small field inflation or large field inflation depending on the value of $\alpha$. If $\alpha\ll 1$, then $C\gg 1$ and
$\Delta\phi$ is small. If $\alpha$ is close to $1/2$, then $C$ is close to zero, and $\Delta\phi$ is large.

\begin{figure*}
$\begin{array}{cc}
\includegraphics[width=0.46\textwidth]{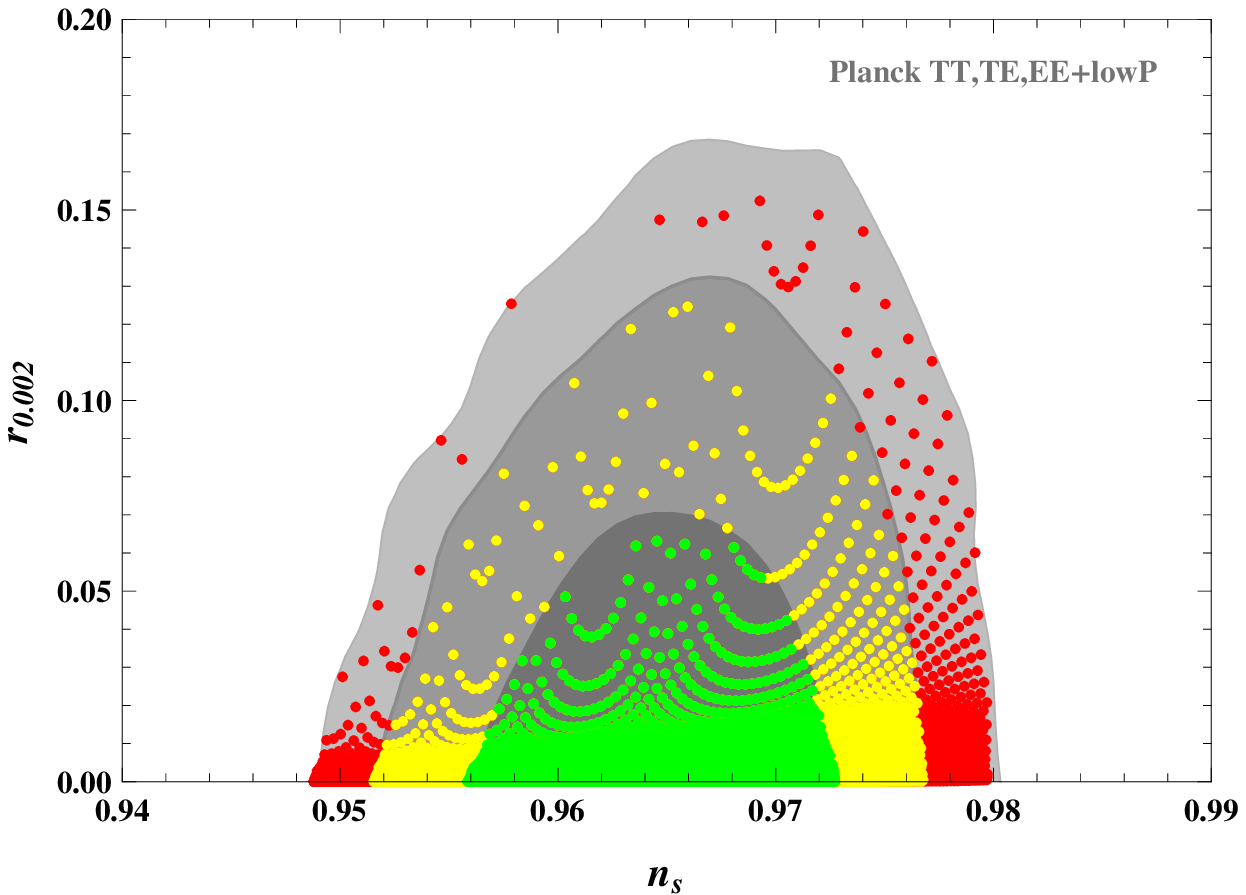}&
\includegraphics[width=0.46\textwidth]{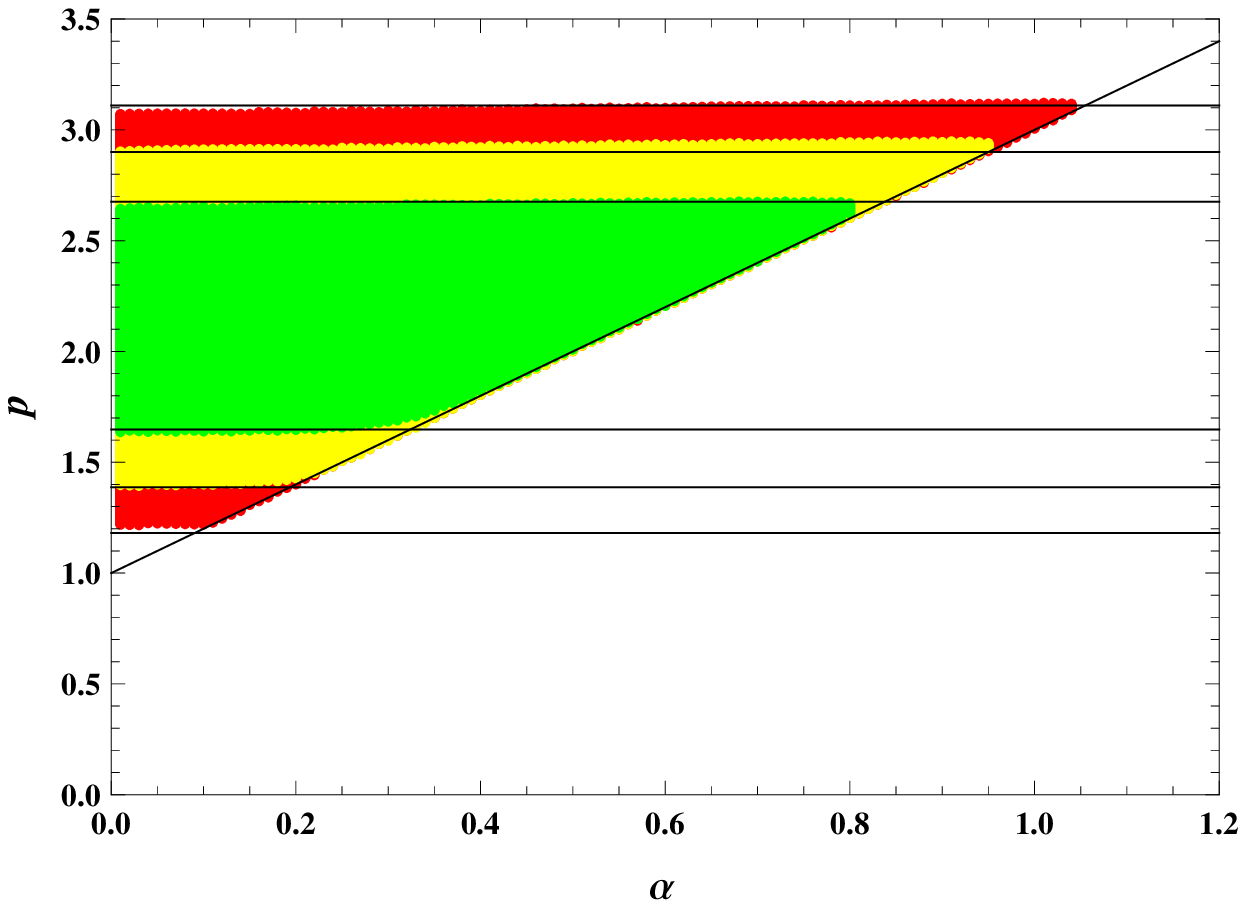}
\end{array}$
\caption{The marginalized 68\%, 95\% and 99.8\% CL contours for $n_s$ and $r_{0.002}$ from Planck 2015 data \citep{Ade:2015lrj} and the observational
constraints on the parametrization \eqref{epsilonneq1}.
The left-hand panel shows the $n_s-r$ contours and the right-hand panel shows the constraints on $p$ and $\alpha$ for $N=60$.
The green, yellow and red regions correspond to 68\%, 95\% and 99.8\% CLs, respectively.}
\label{pc}
\end{figure*}

For the general case with $C>0$ and $p\neq 2$, we get
\begin{equation}
\label{phineq6}
\begin{split}
\phi(N)&=\phi_0\pm \frac{2}{2-p}\sqrt{\frac{2(p-1)}{C}}(N+\alpha)^{1-p/2}\times\\
&\, _2F_1\left[\frac{1}{2},\frac{p-2}{2(p-1)},\frac{4-3p}{2-2p},-\frac{2(N+\alpha)^{1-p}}{C}\right].
\end{split}
\end{equation}
The analytical form of the potential is not apparent, so we analyse the asymptotic form of the potential. For $C\ll 1$, the potential will be the same as the case with $C=0$,
and it is the power-law potential. For $C>1$, Equation (\ref{epsilonneq1}) can be approximated as
\begin{equation}
\label{epsneq7}
\epsilon(N)\approx \frac{p-1}{C(N+\alpha)^p},
\end{equation}
and Equation (\ref{phineq6}) can be approximated as
\begin{equation}
\label{phineq7}
\phi(N)=\phi_0\pm \frac{2}{2-p}\sqrt{\frac{2(p-1)\alpha^p}{p-1-2\alpha}}(N+\alpha)^{(2-p)/2},
\end{equation}
and
\begin{equation}
\label{phineq8}
\phi_e=\phi_0\pm \frac{2\alpha}{2-p}\sqrt{\frac{2(p-1)}{p-1-2\alpha}}.
\end{equation}
Combining Equations (\ref{vpareq1}) and (\ref{phineq7}), for $C>1$, the potential is
\begin{equation}
\label{vphieq6}
\begin{split}
V(\phi)=&\frac{2(p-1)\alpha^p}{A(p-1-2\alpha)}\left\{1\mp \left(\frac{2\alpha^p}{p-1-2\alpha}\right)^{1/(2-p)}\times \right.\\
&\left.\left[\frac{2-p}{2\sqrt{p-1}}(\phi-\phi_0)\right]^{-2(p-1)/(2-p)}\right\}^{-1}.
\end{split}
\end{equation}
If $p>2$, then for small field $\phi$, the potential reduces to the hilltop potential $V(\phi)=V_0[1-(\phi/M)^n]$ \citep{Boubekeur:2005zm} with $n=2(p-1)/(p-2)$ \citep{Garcia-Bellido:2014gna,Creminelli:2014nqa}.
If $1<p<2$, then for large field $\phi$, the potential reduces to the form $V(\phi)=V_0[1-(M/\phi)^n]$ with $n=2(p-1)/(2-p)$ \citep{Garcia-Bellido:2014gna,Creminelli:2014nqa}.
Fitting the model with general $p$ and $\alpha$ to the observational data \citep{Ade:2015lrj}, we find the constraints on $p$ and $\alpha$ for $N=60$
and the results are shown in Fig. \ref{pc}. The results tell us that the model can accommodate both small and large field inflation.
Because $p>2\alpha+1$, so the parametrization (\ref{nsappoxeq1}) requires that $(1-n_s)(N+\alpha)>2\alpha+1$.
At the 99.8\% CL, $0.02\la n_s\la 0.05$, so $\alpha \la 1.03$ and $1.18<p<3.11$ if we take $N=60$.

\section{The parametrization of $\epsilon(N)$}

For the parametrization
\begin{equation}
\label{eplneq41}
\epsilon(N)=\frac{\alpha}{1+s\exp(-\beta N)},
\end{equation}
we get
\begin{equation}
\label{eplneq41a}
n_s-1=\frac{-2\alpha+\beta s \exp(-\beta N)}{1+s\exp(-\beta N)}.
\end{equation}
Note that for the parametrization in this section, we take $s=\pm 1$ for simplicity, $\alpha>0$, $\beta>0$ and the parameter $N=N_*+N_e$ so that we consider
the contribution from $\phi_e$, i.e., at the end of inflation, $0<N=N_e\ll N_*$. When $s=-1$, we require $\exp(\beta N_e)>1$.
Substitute the parametrization \eqref{eplneq41} into Equation \eqref{sclreq2}, we get
\begin{equation}
\label{eplneq42}
V(N)=V_0[s+\exp(\beta N)]^{2\alpha/\beta}.
\end{equation}
Using Equation \eqref{nphieq1}, we get
\begin{equation}
\label{eplneq43}
\phi(N)=\phi_0\pm \frac{2\sqrt{2\alpha}}{\beta}\ln\left[\exp(\beta N/2)+\sqrt{s+\exp(\beta N)}\right],
\end{equation}
and
\begin{equation}
\label{eplneq43a}
\phi_e=\phi_0\pm \frac{2\sqrt{2\alpha}}{\beta}\ln\left[\exp(\beta N_e/2)+\sqrt{s+\exp(\beta N_e)}\right].
\end{equation}
Combining Equations \eqref{eplneq42} and \eqref{eplneq43}, we get
\begin{equation}
\label{eplneq44}
V(\phi)=V_0\left[s+\left(\frac{U(\phi)-sU^{-1}(\phi)}{2}\right)^2\right]^{2\alpha/\beta},
\end{equation}
where $U(\phi)=\exp(\pm \beta(\phi-\phi_0)/(2\sqrt{2\alpha}))$.
For $s=1$, the potential is
\begin{equation}
\label{eplneq44a}
V(\phi)=V_0\left[\cosh\left(\frac{\beta(\phi-\phi_0)}{2\sqrt{2\alpha}}\right)\right]^{4\alpha/\beta}.
\end{equation}
By fitting Equations \eqref{eplneq41} and \eqref{eplneq41a} with $s=1$ to the Planck 2015 data \citep{Ade:2015lrj}, we find that no $\alpha$ and $\beta$ satisfies the 99.8\% CL constraints.

\begin{figure*}
$\begin{array}{cc}
\includegraphics[width=0.46\textwidth]{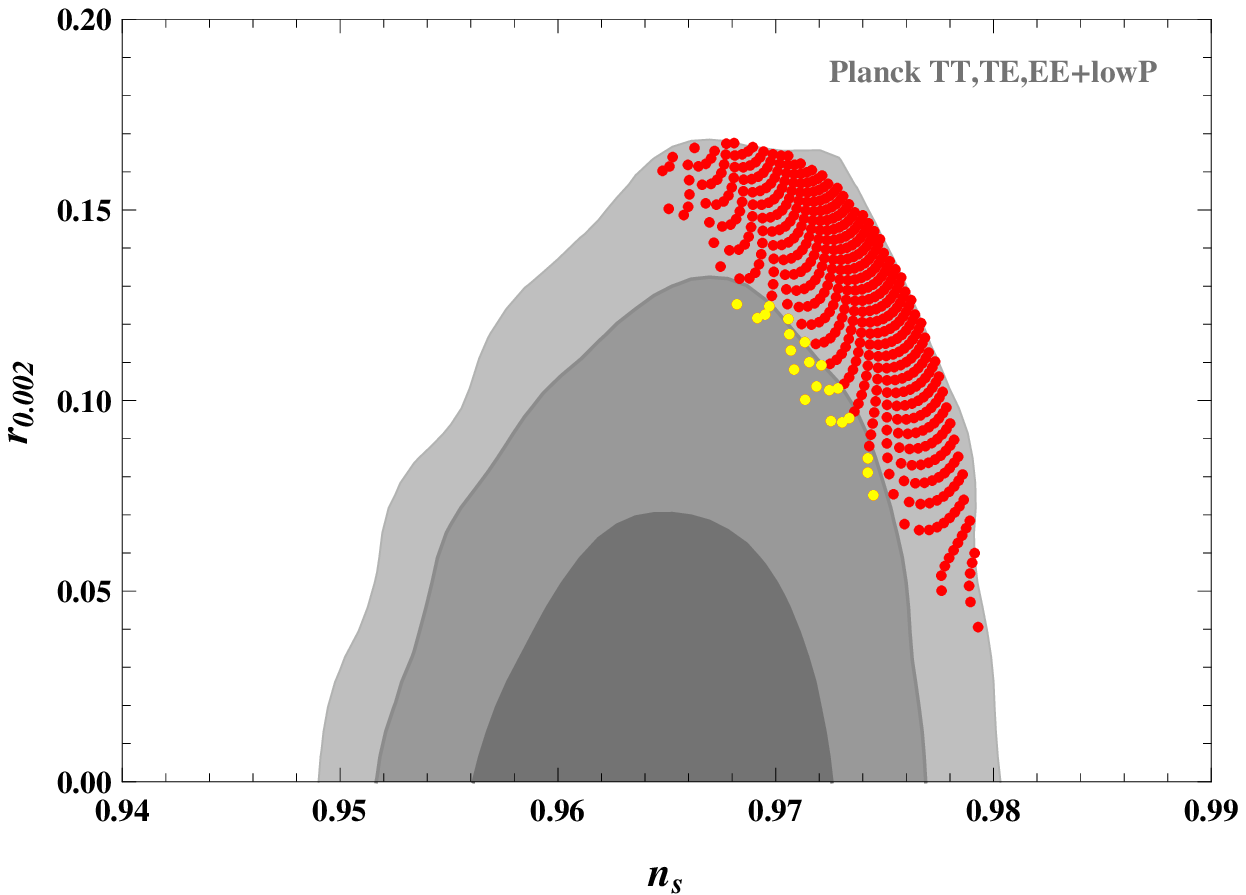}&
\includegraphics[width=0.46\textwidth]{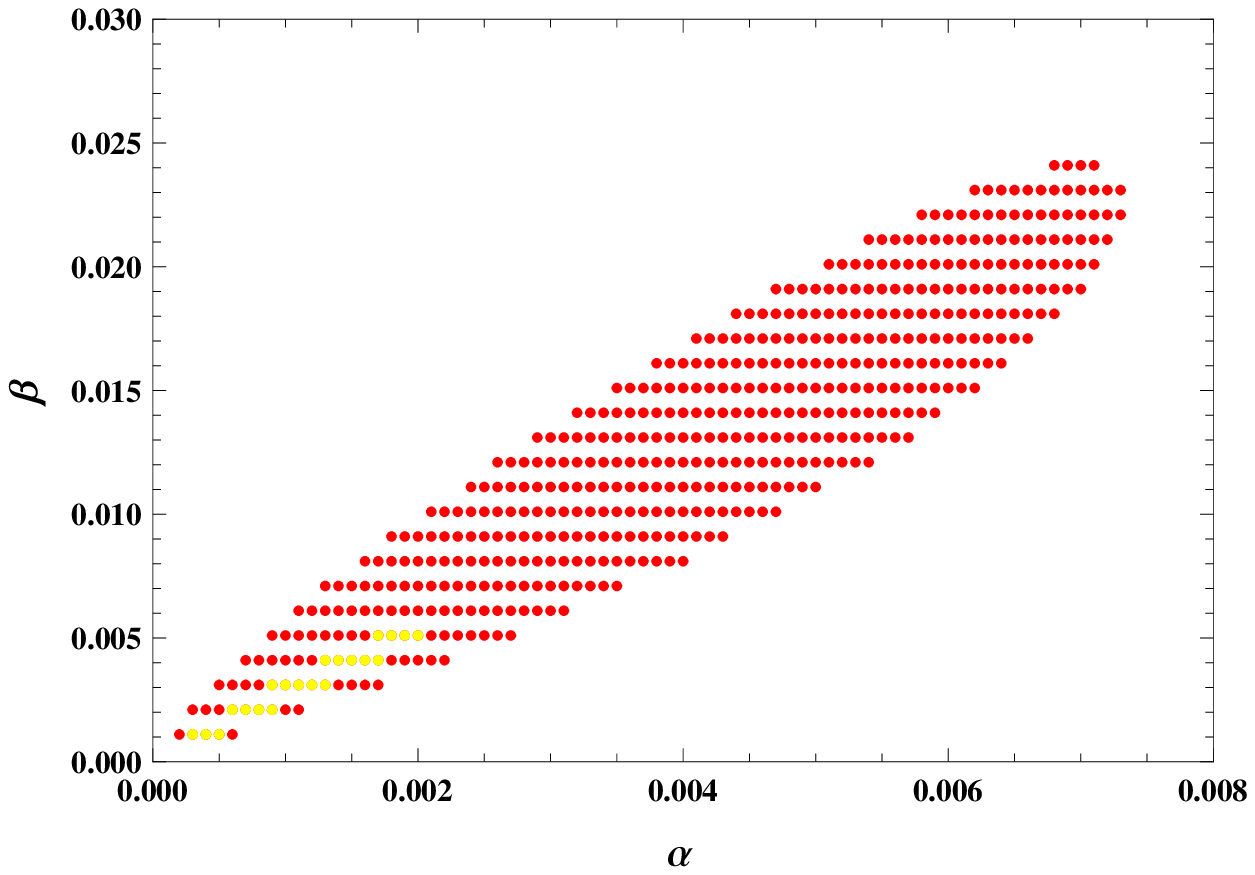}
\end{array}$
\caption{The marginalized 68\%, 95\% and 99.8\% CL contours for $n_s$ and $r_{0.002}$ from Planck 2015 data \citep{Ade:2015lrj} and the observational
constraints on the parametrizations \eqref{eplneq41} with $s=-1$.
The left-hand panel shows the $n_s-r$ contours and the right-hand panel shows the constraints on $\alpha$ and $\beta$ for $N=60$.
The yellow and red regions correspond to 95\% and 99.8\% CLs, respectively.}
\label{eps1}
\end{figure*}

For $s=-1$, the potential is
\begin{equation}
\label{eplneq44b}
V(\phi)=V_0\left[\sinh\left(\frac{\beta(\phi-\phi_0)}{2\sqrt{2\alpha}}\right)\right]^{4\alpha/\beta}.
\end{equation}
Fitting Equations \eqref{eplneq41} and \eqref{eplneq41a} with $s=-1$ to the Planck 2015 data \citep{Ade:2015lrj},
we find that no $\alpha$ and $\beta$ satisfies the 68\% CL constraints, so the model is disfavoured at the 68\% CL.
The 95\% and 99.8\% CL constraints on $\alpha$ and $\beta$ for $N=60$ are shown in Fig. \ref{eps1}.
By using the 99.8\% CL constraints, we find that $8.37\leq \Delta\phi \leq 16.48$, so the field
excursion of the inflaton in this model is super-Planckian.

\begin{figure*}
$\begin{array}{cc}
\includegraphics[width=0.46\textwidth]{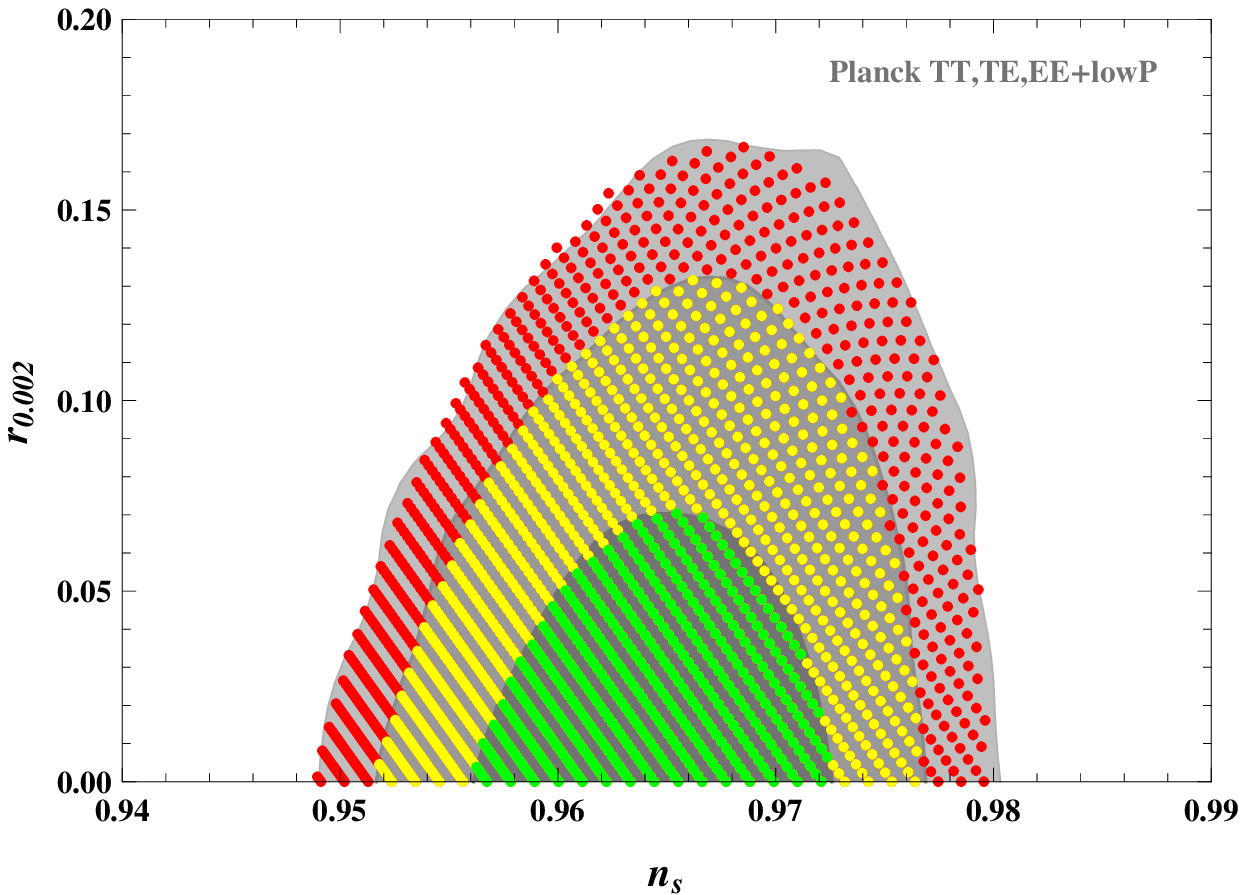}&
\includegraphics[width=0.46\textwidth]{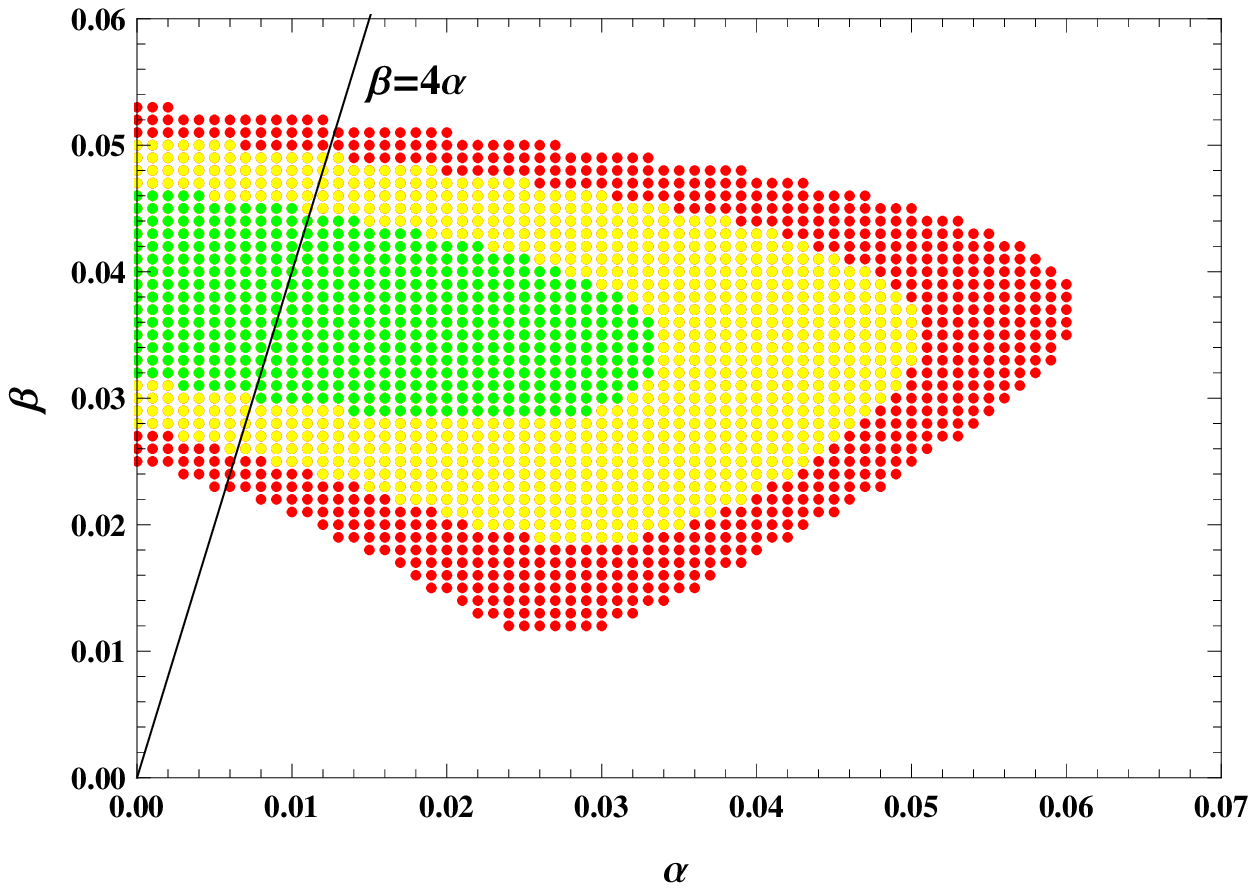}
\end{array}$
\caption{The marginalized 68\%, 95\% and 99.8\% CL contours for $n_s$ and $r_{0.002}$ from Planck 2015 data \citep{Ade:2015lrj} and the observational
constraints on the parametrizations \eqref{eplneq45} with $s=1$.
The left-hand panel shows the $n_s-r$ contours and the right-hand panel shows the constraints on $\alpha$ and $\beta$ for $N=60$.
The green, yellow and red regions correspond to 68\%, 95\% and 99.8\% CLs, respectively. The solid line $\beta=4\alpha$ in the right-hand panel corresponds
to the s-dual inflation.}
\label{eps2}
\end{figure*}

For the parametrization
\begin{equation}
\label{eplneq45}
\epsilon(N)=\frac{\alpha\exp(-\beta N)}{1+s\exp(-\beta N)},
\end{equation}
we get
\begin{equation}
\label{eplneq46}
n_s-1=-\frac{\beta+2\alpha\exp(-\beta N)}{1+s\exp(-\beta N)}.
\end{equation}
Substitute the parametrization \eqref{eplneq45} into Equation \eqref{sclreq2}, we get
\begin{equation}
\label{eplneq47}
V(N)=V_0[1+s\exp(-\beta N)]^{-2\alpha/(\beta s)}.
\end{equation}
From Equation \eqref{nphieq1}, we get
\begin{equation}
\label{eplneq48b}
\phi(N)=\phi_0\mp \frac{2\sqrt{2\alpha}}{\beta\sqrt{s}}{\rm arcoth}\left[\sqrt{\frac{s+\exp(\beta N)}{s}}\,\right].
\end{equation}
For $s=1$, we get
\begin{equation}
\label{eplneq48c}
\phi_e=\phi_0\mp \frac{2\sqrt{2\alpha}}{\beta}{\rm arcoth}\left[\sqrt{1+\exp(\beta N_e})\,\right],
\end{equation}
and the potential
\begin{equation}
\label{eplneq49a}
V(\phi)=V_0\left[{\rm sech}\left(\frac{\beta(\phi-\phi_0)}{2\sqrt{2\alpha}}\right)\right]^{4\alpha/\beta}.
\end{equation}
If $\beta=4\alpha=2/M^2$, we recover the potential for the s-dual inflation $V=V_0{\rm sech}(\phi/M)$ \citep{Anchordoqui:2014uua}.
Fitting Equations \eqref{eplneq45} and \eqref{eplneq46} with $s=1$ to the Planck 2015 data \citep{Ade:2015lrj},
we obtain the constraints on the parameters $\alpha$ and $\beta$ for $N=60$ and the results are shown in Fig. \ref{eps2}.
From Fig. \ref{eps2}, we see that the s-dual inflation is consistent with the observational data.
By using the 99.8\% CL constraints, we find that $0<\Delta\phi \leq 26.57$, so the model includes
both the large field and small field inflation. If $\alpha$ is close to zero, then $\Delta\phi$ is small.

\begin{figure*}
$\begin{array}{cc}
\includegraphics[width=0.46\textwidth]{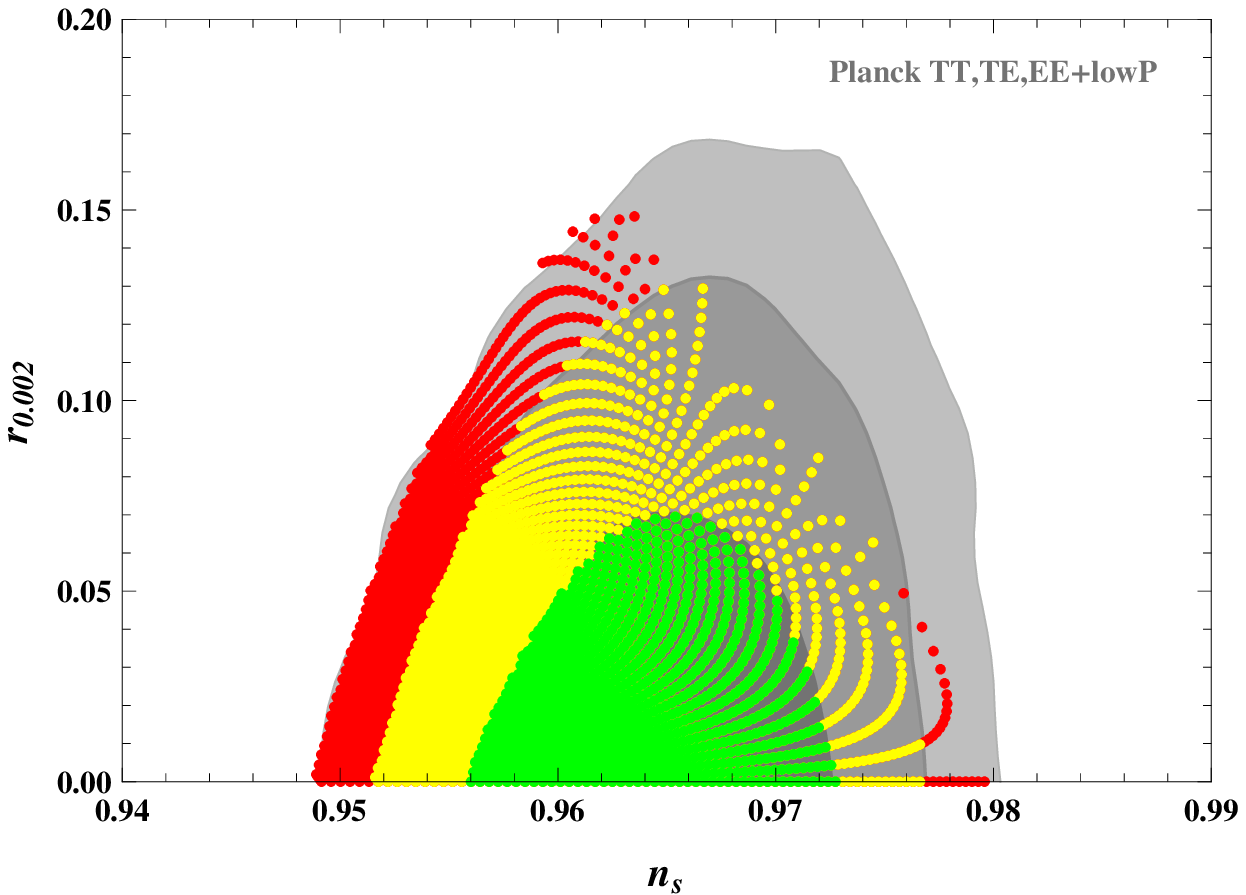}&
\includegraphics[width=0.46\textwidth]{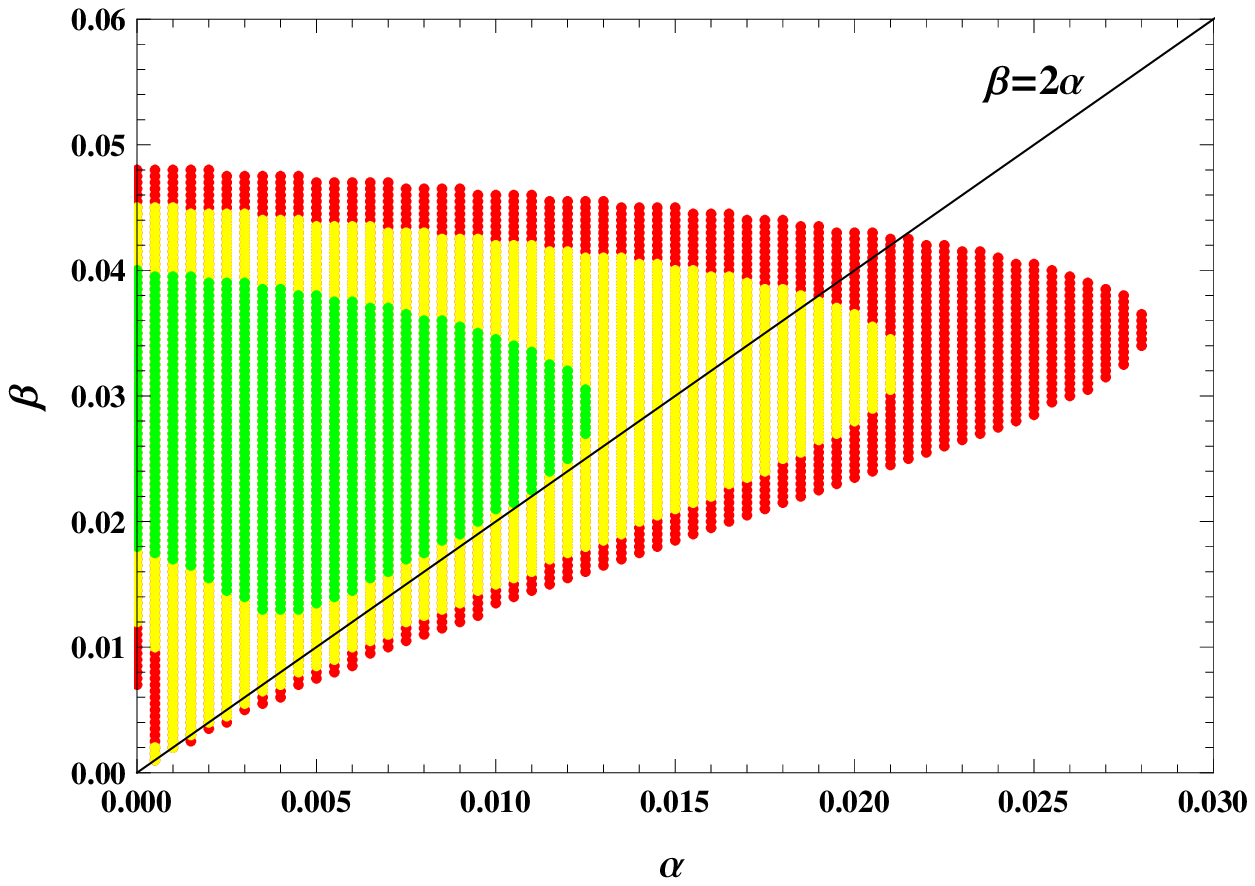}
\end{array}$
\caption{The marginalized 68\%, 95\% and 99.8\% CL contours for $n_s$ and $r_{0.002}$ from Planck 2015 data \citep{Ade:2015lrj} and the observational
constraints on the parametrizations \eqref{eplneq45} with $s=-1$.
The left-hand panel shows the $n_s-r$ contours and the right-hand panel shows the constraints on $\alpha$ and $\beta$ for $N=60$.
The green, yellow and red regions correspond to 68\%, 95\% and 99.8\% CLs, respectively.
The solid line $\beta=2\alpha$ in the right-hand panel corresponds
to the natural inflation.}
\label{eps3}
\end{figure*}

For $s=-1$, we get the potential,
\begin{equation}
\label{eplneq410}
V(\phi)=V_0\left[\sin\left(\frac{\beta(\phi-\phi_0)}{2\sqrt{2\alpha}}\right)\right]^{4\alpha/\beta},
\end{equation}
or
\begin{equation}
\label{eplneq410a}
\begin{split}
V(\phi)&=V_0\left[\cos\left(\frac{\beta(\phi-\phi_0)}{2\sqrt{2\alpha}}\right)\right]^{4\alpha/\beta}\\
&=\frac{V_0}{4}\left[1+\cos\left(\frac{\beta(\phi-\phi_0)}{\sqrt{2\alpha}}\right)\right]^{2\alpha/\beta},
\end{split}
\end{equation}
If we take $\beta=2\alpha$ and $f=1/\sqrt{2\alpha}$, then we recover the potential for natural inflation $V(\phi)=\Lambda^4[1+\cos(\phi/f)]$ \citep{Freese:1990rb}.
Fitting Equations \eqref{eplneq45} and  \eqref{eplneq46} with $s=-1$ to the Planck 2015 data \citep{Ade:2015lrj},
we obtain the constraints on the parameters $\alpha$ and $\beta$ for $N=60$ and the results are shown in Fig. \ref{eps3}.
From Fig. \ref{eps3}, we see that natural inflation is disfavoured at the 68\% CL.
By using the 99.8\% CL constraints, we find that $0<\Delta\phi \leq 54.41$, so the model includes
both the large field and small field inflation, and the small field inflation is achieved when $\alpha$ is close to zero.

\begin{figure*}
$\begin{array}{cc}
\includegraphics[width=0.46\textwidth]{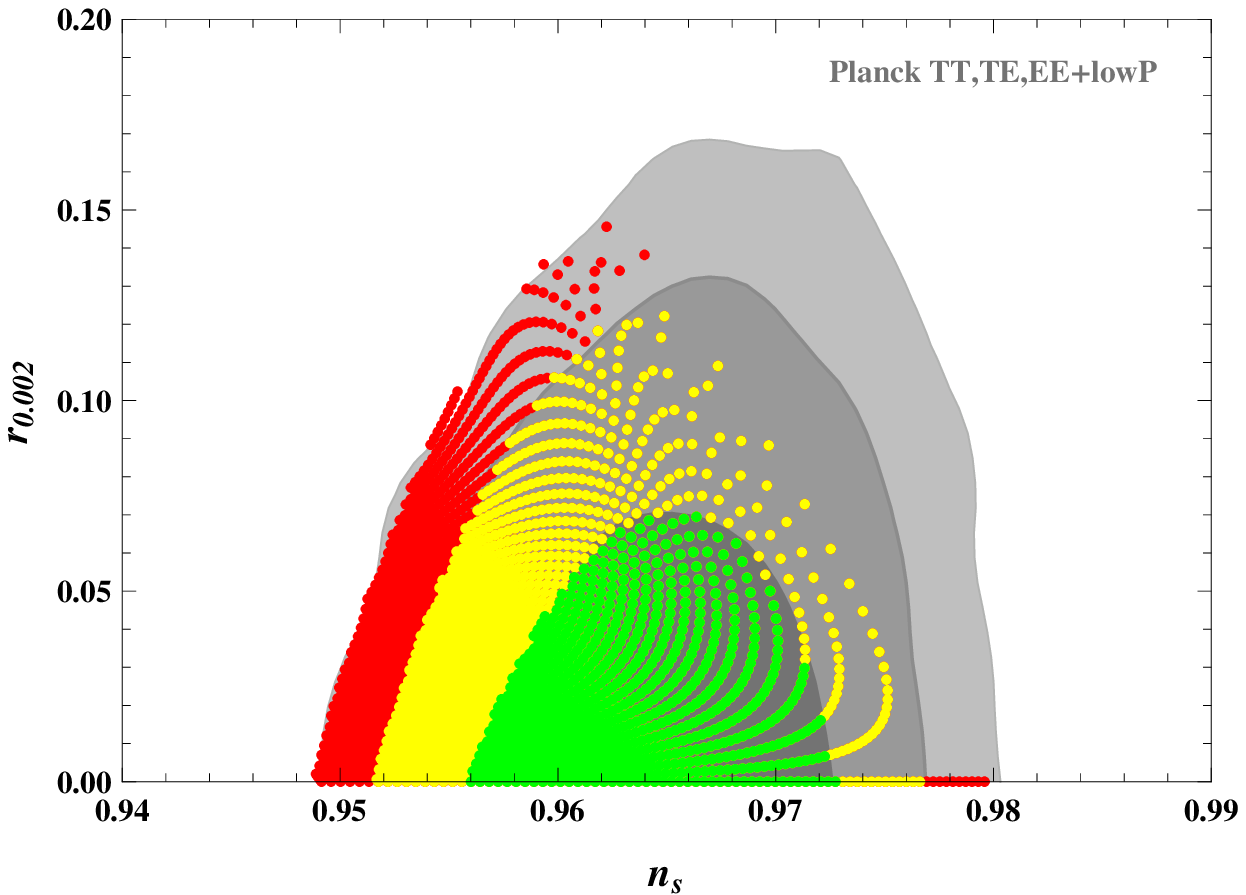}&
\includegraphics[width=0.46\textwidth]{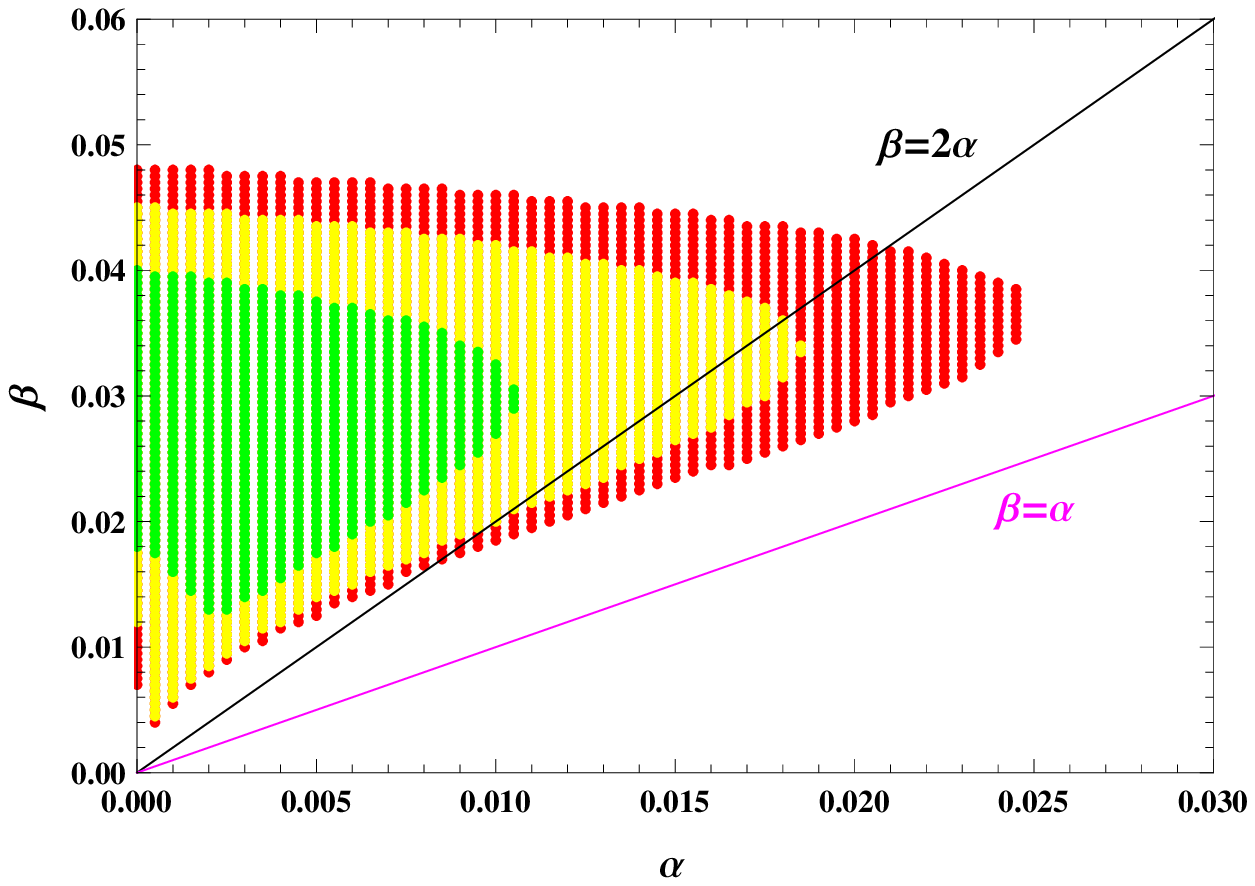}
\end{array}$
\caption{The marginalized 68\%, 95\% and 99.8\% CL contours for $n_s$ and $r_{0.002}$ from Planck 2015 data \citep{Ade:2015lrj} and the observational
constraints on the parametrizations \eqref{eplneq411} with $s=-1$.
The left-hand panel shows the $n_s-r$ contours and the right-hand panel shows the constraints on $\alpha$ and $\beta$ for $N=60$.
The green, yellow and red regions correspond to 68\%, 95\% and 99.8\% CLs, respectively.
The black solid line $\beta=2\alpha$ in the right-hand panel corresponds
to the hilltop potential with $p=2$, and the magenta solid line $\beta=\alpha$ corresponds to the double well potential.}
\label{eps4}
\end{figure*}

For the parametrization
\begin{equation}
\label{eplneq411}
\epsilon(N)=\frac{\alpha\exp(-\beta N)}{[1+s\exp(-\beta N)]^2},
\end{equation}
we get
\begin{equation}
\label{eplneq412}
n_s-1=\frac{-\beta-2\alpha\exp(-\beta N)+\beta s^2\exp(-2\beta N)}{[1+s\exp(-\beta N)]^2},
\end{equation}
\begin{equation}
\label{eplneq413}
V(N)=V_0\exp\left[\frac{-2\alpha}{\beta[s+\exp(\beta N)]}\right],
\end{equation}
and
\begin{equation}
\label{eplneq414}
\phi(N)=\phi_0\pm \frac{2\sqrt{2\alpha}}{\beta\sqrt{s}}\arctan\left[\exp(\beta N/2)/\sqrt{s}\,\right].
\end{equation}
For $s=-1$, we get the potential
\begin{equation}
\label{eplneq415}
V(\phi)=V_0\exp\left[-\frac{2\alpha}{\beta}\sinh^2\left(\frac{\beta(\phi-\phi_0)}{2\sqrt{2\alpha}}\right)\right].
\end{equation}
If $\beta=2\alpha=4/\mu^2$, then the potential reduces to the hilltop potential $V(\phi)=V_0[1-(\phi/\mu)^p]$ with $p=2$ for small $\phi$.
If $\alpha=\beta=8/\mu^2$, then the potential reduces to the double well potential $V(\phi)=V_0[1-(\phi/\mu)^2]^2$ for small $\phi$ \citep{Olive:1989nu}.
Fitting Equations \eqref{eplneq411} and \eqref{eplneq412} with $s=-1$ to the Planck 2015 data \citep{Ade:2015lrj},
we obtain the constraints on the parameters $\alpha$ and $\beta$ for $N=60$ and the results are shown in Fig. \ref{eps4}.
From Fig. \ref{eps4}, we see that the double well potential is excluded by the observational data
and the hilltop potential with $p=2$ is disfavoured at the 68\% CL.
By using the 99.8\% CL constraints, we find that $0<\Delta\phi \leq 14.31$, so the model includes
both the large field and small field inflation. If $\alpha$ is close to zero, then $\Delta\phi$ is small.

\begin{figure*}
$\begin{array}{cc}
\includegraphics[width=0.46\textwidth]{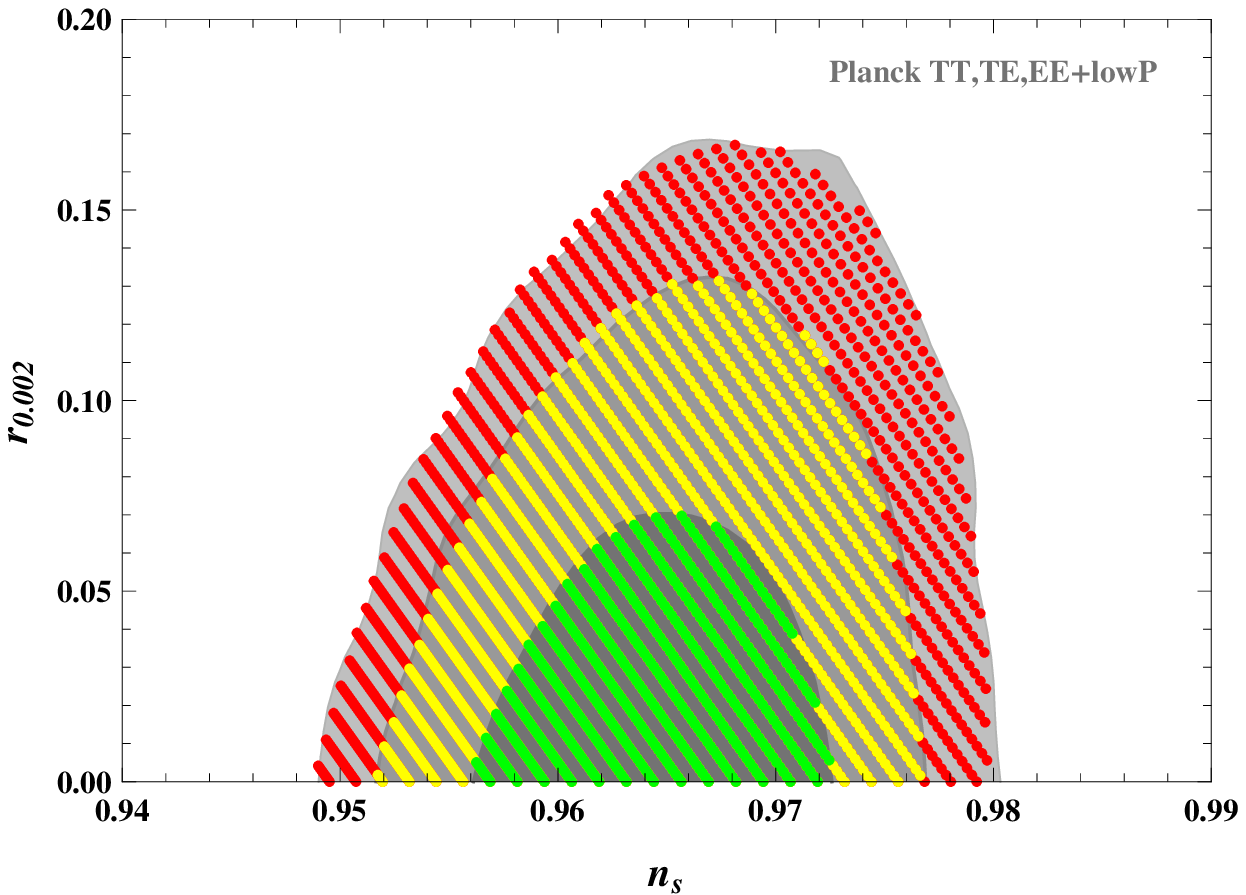}&
\includegraphics[width=0.46\textwidth]{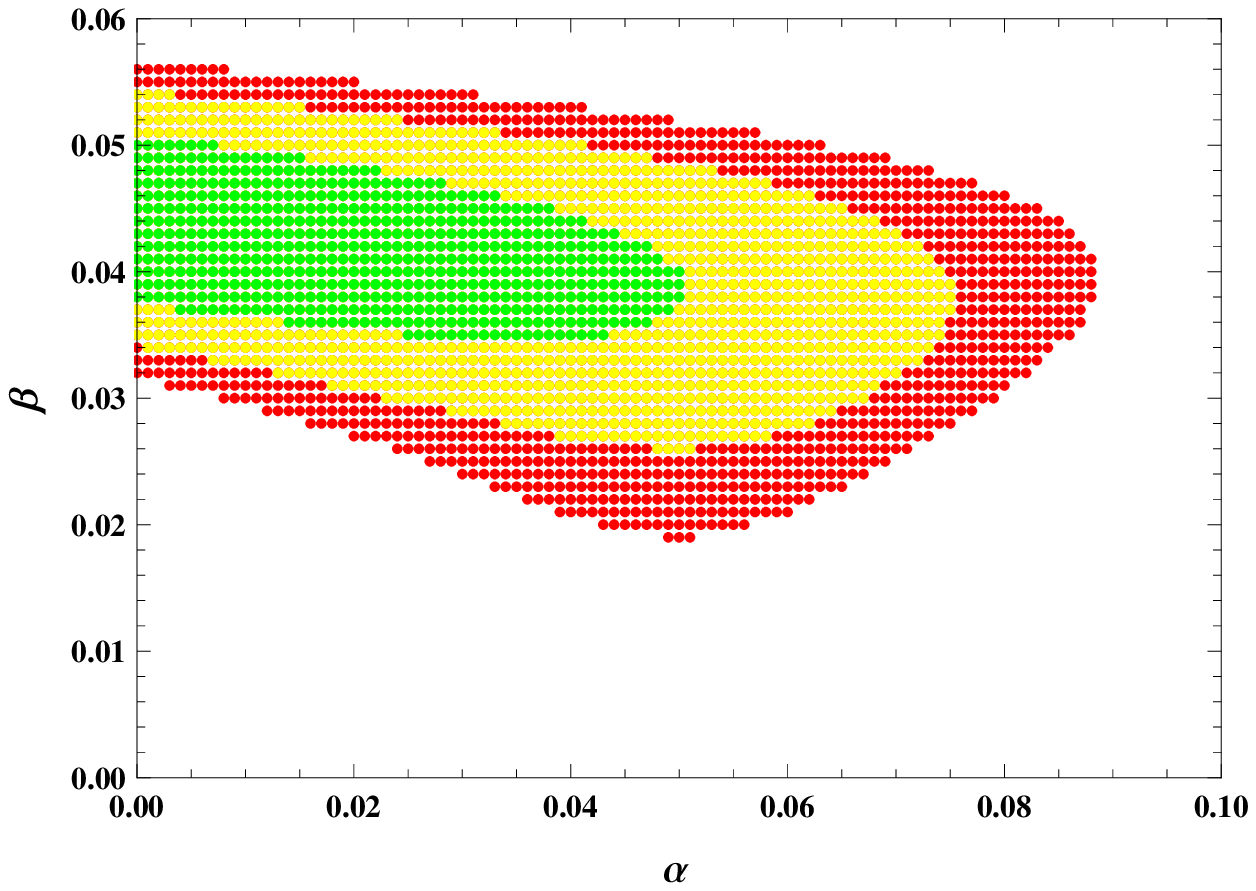}
\end{array}$
\caption{The marginalized 68\%, 95\% and 99.8\% CL contours for $n_s$ and $r_{0.002}$ from Planck 2015 data \citep{Ade:2015lrj} and the observational
constraints on the parametrizations \eqref{eplneq411} with $s=1$.
The left-hand panel shows the $n_s-r$ contours and the right-hand panel shows the constraints on $\alpha$ and $\beta$ for $N=60$.
The green, yellow and red regions correspond to 68\%, 95\% and 99.8\% CLs, respectively.}
\label{eps5}
\end{figure*}

For $s=1$, we get the potential
\begin{equation}
\label{eplneq416}
V(\phi)=V_0\exp\left[-\frac{\alpha}{\beta}\left(1+\cos\frac{\beta(\phi-\phi_0)}{\sqrt{2\alpha}}\right)\right].
\end{equation}
Fitting Equations \eqref{eplneq411} and \eqref{eplneq412} with $s=1$ to the Planck 2015 data \citep{Ade:2015lrj},
we obtain the constraints on the parameters $\alpha$ and $\beta$ for $N=60$ and the results are shown in Fig. \ref{eps5}.
By using the 99.8\% CL constraints, we find that $0<\Delta\phi \leq 31.23$, so the model includes
both the large field and small field inflation, and the small field inflation is achieved if $\alpha$ is close to zero.

\section{The parametrization of $\phi(N)$}

Combining Equations (\ref{sclreq2}) and (\ref{nphieq1}), we get
\begin{equation}
\label{vphineq4}
(\ln V)_{,N}=(\phi_{,N})^2.
\end{equation}
Once the functional form $\phi(N)$ is known, we can derive the potential form $V(\phi)$.
Let us first consider the power-law parametrization
\begin{equation}
\label{phinpareq1}
\phi(N)=\sigma(N+\gamma)^{\beta}.
\end{equation}
For $\beta=1/2$, from Equation \eqref{vphineq4}, we get the power-law potential,
\begin{equation}
V(\phi)=V_0(N+\gamma)^{\frac{\sigma^2}{4}}=V_0\left(\frac{\phi}{\sigma}\right)^{\frac{\sigma^2}{2}},
\end{equation}
where $V_0$ is and integration constant. From Equation \eqref{nphieq1}, we get
\begin{equation}
\label{phingq}
\epsilon=\frac{\sigma^2}{8(N+\gamma)},
\end{equation}
so
\begin{equation}
\label{phingq1}
n_s-1=-\frac{1+\sigma^2/4}{N+\gamma}, \quad r=\frac{2\sigma^2}{N+\gamma},
\end{equation}
and $\sigma^2\approx 8\gamma$. The results are the same as those discussed in Section 3 with $C=0$, $p=1+\sigma^2/4$ and $\alpha=\gamma$.

\begin{figure*}
$\begin{array}{cc}
\includegraphics[width=0.46\textwidth]{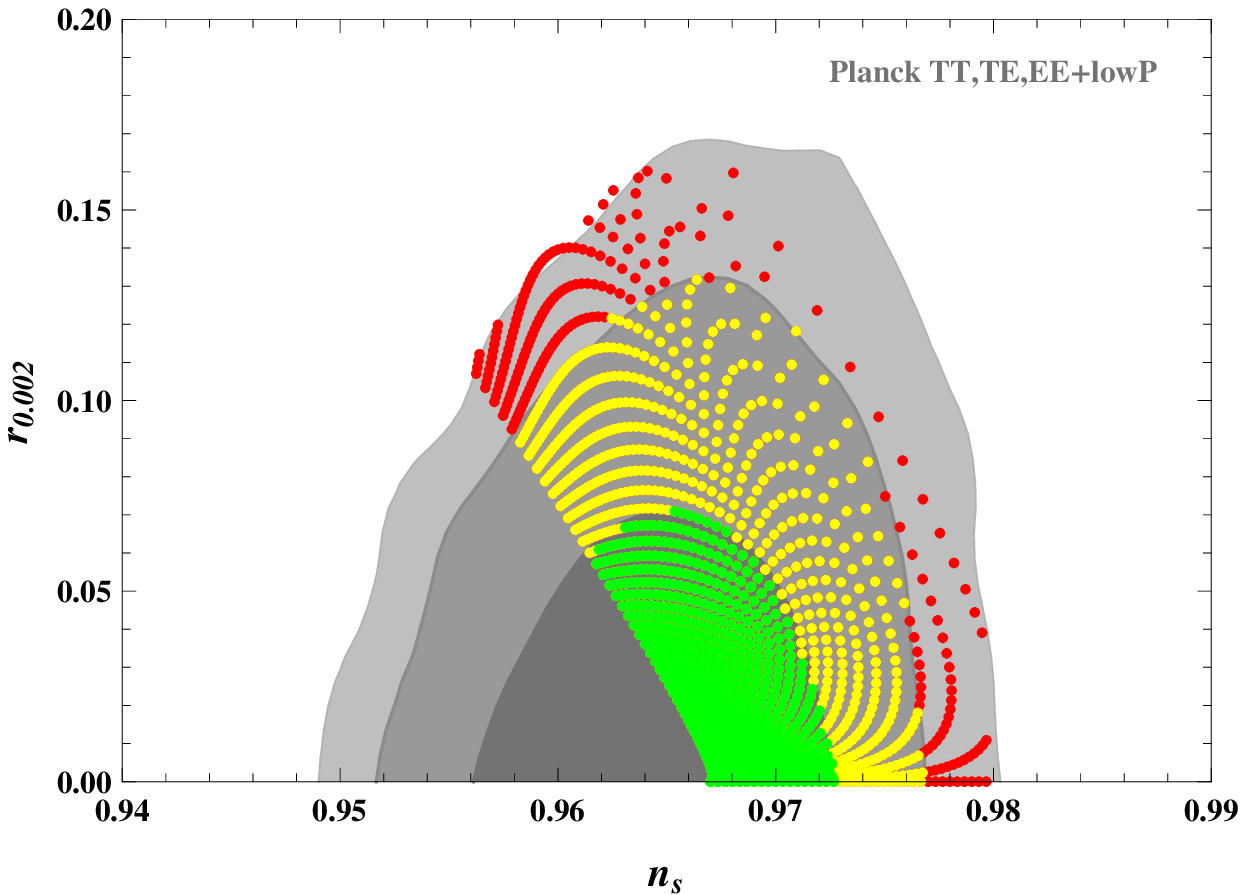}&
\includegraphics[width=0.46\textwidth]{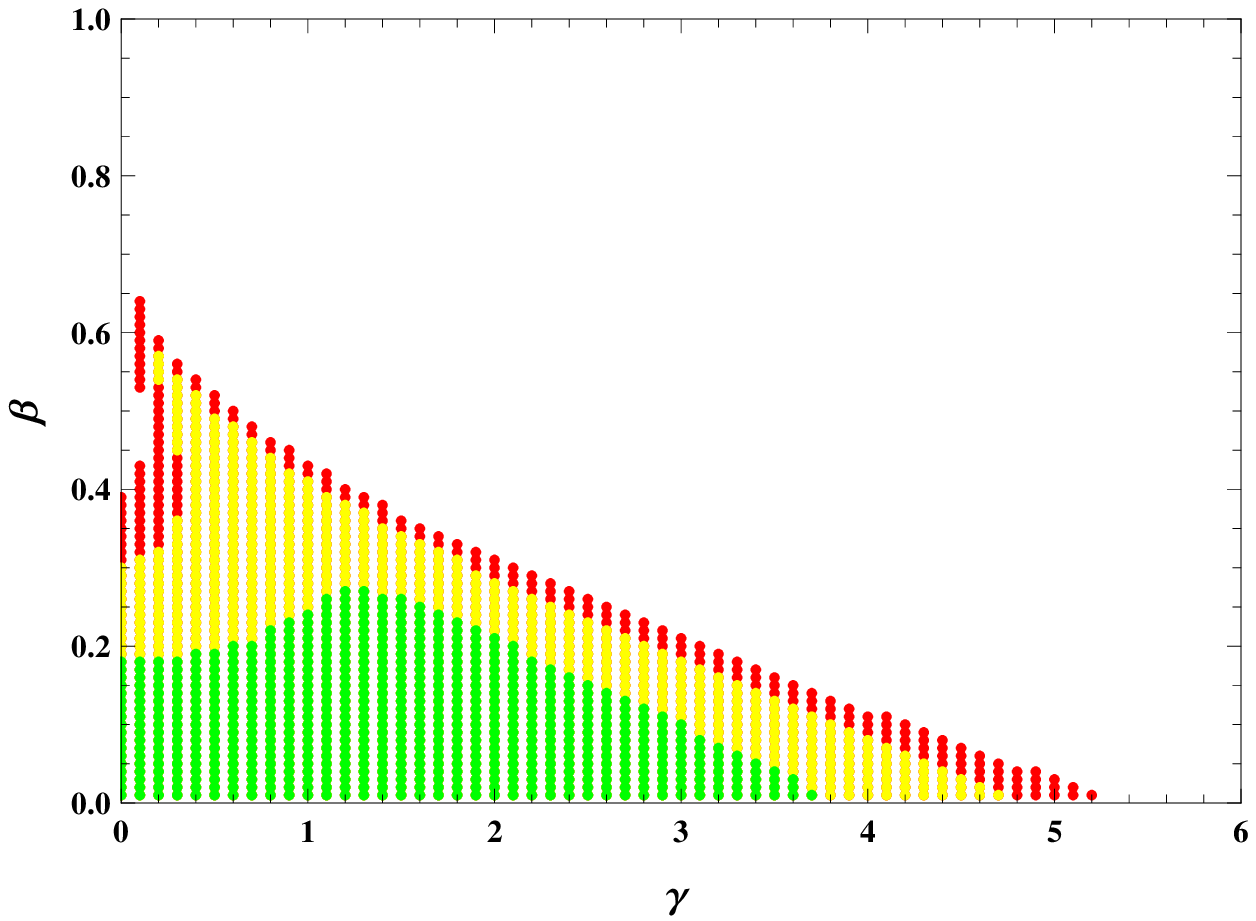}
\end{array}$
\caption{The marginalized 68\%, 95\% and 99.8\% CL contours for $n_s$ and $r_{0.002}$ from Planck 2015 data \citep{Ade:2015lrj} and the observational
constraints on the parametrizations \eqref{phinpareq1}.
The left-hand panel shows the $n_s-r$ contours and the right-hand panel shows the constraints on $\beta$ and $\gamma$ for $N=60$.
The green, yellow and red regions correspond to 68\%, 95\% and 99.8\% CLs, respectively.}
\label{eps6}
\end{figure*}

For $\beta\neq 1/2$, we derive the potential,
\begin{equation}
\label{phinveq29}
\begin{split}
V(\phi)&=V_0\exp\left[\frac{\sigma^2\beta^2}{2\beta-1}(N+\gamma)^{2\beta-1}\right]\\
&=V_0\exp\left[\frac{\sigma^2\beta^2}{2\beta-1}\left(\frac{\phi}{\sigma}\right)^{2-1/\beta}\right],
\end{split}
\end{equation}
the scalar spectral tilt,
\begin{equation}
\label{phinnseq30}
n_s-1=\frac{2\beta-2}{N+\gamma}-\sigma^2\beta^2(N+\gamma)^{2\beta-2},
\end{equation}
the tensor to scalar ratio,
\begin{equation}
\label{phinreq31}
r=8\sigma^2\beta^2(N+\gamma)^{2\beta-2},
\end{equation}
and the parameters $\sigma$, $\beta$ and $\gamma$ satisfy the relation $\sigma^2\beta^2\approx 2\gamma^{2-2\beta}$.
Note that Equations \eqref{phinnseq30} and \eqref{phinreq31} include the special case $\beta=1/2$ which corresponds to the power-law potential.
Fitting Equations \eqref{phinnseq30} and \eqref{phinreq31} for the parametrization \eqref{phinpareq1}  to the Planck 2015 data \citep{Ade:2015lrj},
we obtain the constraints on the parameters $\alpha$ and $\beta$ for $N=60$ and the results are shown in Fig. \ref{eps6}.
By using the 99.8\% CL constraints, we find that $0<\Delta\phi \leq 754.23$.
If $\beta$ is near zero, then it is large field inflation. If $\gamma$
is near zero, then it is small field inflation. In the left-hand panel of Fig. \ref{eps6}, the lower bounds of the $n_s-r$ contours are set by $\beta=0$.

Next we consider the parametrization $\phi(N)=\sigma\ln(\beta N+\gamma)$. Combining Equations \eqref{nsapproxeq2}, \eqref{nsapproxeq4}, \eqref{nphieq1} and
\eqref{vphineq4}, we get
\begin{gather}
\label{phinvphieq32}
\begin{split}
V(\phi)&=V_0\exp\left(-\frac{\sigma^2\beta}{\beta N+\gamma}\right)\\
&=V_0\exp\left[-\sigma^2\beta\exp(-\phi/\sigma)\right],
\end{split}\\
\label{phinepsiloneq33}
\epsilon(N)=\frac{\sigma^2\beta^2}{2(\beta N+\gamma)^2},\\
n_s-1=-\frac{2}{N+\gamma/\beta}-\frac{\sigma^2}{(N+\gamma/\beta)^2},\\
r=\frac{8\sigma^2}{(N+\gamma/\beta)^2}.
\end{gather}
From the end of inflation condition $\epsilon(N=0)\approx 1$, we get $\sigma^2\beta^2=2\gamma^2$, so $r=16(\gamma/\beta)^2/(N+\gamma/\beta)^2$,
and $n_s$ and $r$ depend on the parameter $\gamma/\beta$ only.
If $\gamma/\beta$ is small, then $n_s-1\sim -2/N$ and $r\sim 1/N^2$, and the model will behave like the model \eqref{nsappoxeq1} with $p=2$
and small $\alpha=\gamma/\beta$, so the two models will cover some common regions in the $n_s-r$ graph. The fitting results are shown in Fig. \ref{nsrpower}.
In particular, for $\phi/\sigma\gg 1$, the potential can be approximated as
\begin{equation}
\label{phinvphieq33}
\begin{split}
V(\phi)&\approx V_0\left[1-2\gamma^2 \exp(-\phi/\sigma)\right]\\
&\approx V_0\left[1-\gamma^2 \exp(-\phi/\sigma)\right]^2.
\end{split}
\end{equation}
Therefore, the $R^2$ inflation is also included in this model.

In the last we consider the exponential parametrization $\phi(N)=\sigma\exp(\beta N+\gamma)$ with $\beta<0$. Following the same
procedure as above, we obtain
\begin{gather}
V(\phi)=V_0\exp(\beta\phi^2/2), \\
\epsilon=\frac{\sigma^2\beta^2}{2}\exp(2\beta N+2\gamma)=r/16, \\
n_s-1=2\beta-\beta^2\sigma^2\exp(2\beta N+2\gamma).
\end{gather}
The model parameters satisfy the relation $\beta^2\sigma^2\exp(2\gamma)=2$, so $\epsilon=\exp(\beta N)$ and
$n_s-1=2\beta-2\exp(2\beta N)$. Since $\epsilon<1$, the parameter $\beta$ should be negative.
From the constraint on $r$, we can get the upper limit on $\beta$, with this upper limit, we find that the
model is not consistent with the observational data.

\section{Conclusions}
For the double well potential $V(\phi)=V_0[1-(\phi/\mu)^2]^2$, the potential $V(\phi)=V_0\exp(\beta\phi^2)$
and the potential $V(\phi)=V_0[\cosh(\beta\phi/2\sqrt{2\alpha})]^{4\alpha/\beta}$,
the predicted $n_s$ and $r$ are not consistent with the Planck 2015 data.
The power-law potential $V(\phi)=V_0\phi^{2p-2}$, the potential $V(\phi)=V_0[\sinh(\beta\phi/2\sqrt{2\alpha})]^{4\alpha/\beta}$,
the natural inflation and the hilltop potential with $n=2$ are disfavoured by the observational
data at the 68\% CL. At the 99.8\% CL, we find that $1.2\le p \le 2.21$ for the power-law potential if we take the number of e-folds before the end
of inflation $N=60$.
For the power-law potential $V(\phi)=V_0(\phi-\phi_0)^{2(p-1)}$, the T-model potential $V(\phi)=V_0\tanh^2(\gamma\phi)$ which includes the $\alpha$-attractors
and the Starobinsky model, the hilltop potential $V(\phi)=V_0[1-(\phi/M)^n]$ with $n=2(p-1)/(p-2)$ and $p>2$,
the potential $V(\phi)=V_0[1-(M/\phi)^n]$ with $n=2(p-1)/(2-p)$ and $1<p<2$,
and the potential $V(\phi)=V_0\exp[-\sigma^2\beta\exp(-\phi/\sigma)]$,
their spectral tilts have the universal behavior $n_s=-p/(N+\alpha)$.

For the parametrization $\epsilon(N)=\alpha\exp(-\beta N)/[1+\exp(-\beta N)]$, we get $n_s-1=-[\beta+2\alpha\exp(-\beta N)]/[1+\exp(-\beta N)]$
and the corresponding potential $V(\phi)=V_0[{\rm sech}(\beta\phi/2\sqrt{2\alpha})]^{4\alpha/\beta}$ which includes the s-dual inflation.
For the parametrization $\epsilon(N)=\alpha\exp(-\beta N)/[1-\exp(-\beta N)]$, we get $n_s-1=-[\beta+2\alpha\exp(-\beta N)]/[1-\exp(-\beta N)]$
and the corresponding potential $V(\phi)=V_0[\sin(\beta\phi/2\sqrt{2\alpha})]^{4\alpha/\beta}$ which includes the natural inflation.
For the parametrization $\epsilon(N)=\alpha\exp(-\beta N)/[1-\exp(-\beta N)]^2$, we get the corresponding potential
$V(\phi)=V_0\exp[-2\alpha\sinh^2(\beta\phi/2\sqrt{2\alpha})/\beta]$ which includes the hilltop potential with $n=2$ and the double well potential.
The tensor to scalar ratio $r$ for these models can easily be small due to the factor $\exp(-\beta N)$ in $\epsilon(N)$.
For the parametrization $\phi(N)=\sigma\ln(\beta N+\gamma)$, the corresponding
potential is $V(\phi)=V_0\exp[-\sigma^2\beta\exp(-\phi/\sigma)]$,
both $n_s$ and $r$ depend on $\gamma/\beta$ only and the model has the universal behavior $n_s-1=-2/(N+\gamma/\beta)$ if $\gamma/\beta$ is small.
All these models can achieve both small and large field inflation.

Based on the slow-roll relations (\ref{nsapproxeq4}), (\ref{sclreq2}), (\ref{nsapproxeq6}) and (\ref{nphieq1}), by parameterizing one of the parameters $n_s(N)$,
$\epsilon(N)$ and $\phi(N)$, and fitting the parameters in the models to the observational data, we not only
obtain the constraints on the parameters, but also easily reconstruct the classes
of the inflationary models which include the chaotic inflation, T-model, hilltop inflation,
s-dual inflation, natural inflation and $R^2$ inflation, and the reconstructed inflationary models are consistent with the observations.
Since the observational data only probes a rather small intervals of scales, the reconstructed potentials
approximate the inflationary potential only in the slow-roll regime for the observational scales $10^{-3}~\text{Mpc}\la k^{-1}\la 10^4~\text{Mpc}$.
Outside the slow-roll regime, the inflationary potential can be rather different, but it does not mean that the reconstructed potential is not applicable
in that regime. Once the potential is obtained, we can either apply the slow-roll formulae or work out the exact solutions.

\section*{acknowledgements}

This research was supported in part by the Natural Science
Foundation of China under Grants Nos. 11175270 and 11475065,
and the Program for New Century Excellent Talents in University under Grant No. NCET-12-0205.



\end{document}